\documentclass[useAMS,usenatbib,twocolumn,preprintnumbers,nofootinbib]{revtex4}
\def \be {\begin{equation}} 
\def \ee {\end{equation}} 
\def \bea {\begin{eqnarray}} 
\def \eea {\end{eqnarray}} 

\usepackage{graphicx}
\usepackage{dcolumn}
\usepackage{bm}
\usepackage{epsfig} 
\usepackage{amsfonts}
\usepackage{amsmath}
\usepackage{amssymb}
\usepackage[usenames]{color}

\newcommand*{\ltsim}{\ {\raise-.75ex\hbox{$\buildrel<\over\sim$}}\ }
\newcommand*{\gtsim}{\ {\raise-.75ex\hbox{$\buildrel>\over\sim$}}\ }
\newcommand*{\proptosim}{\ {\raise-.75ex\hbox{$\buildrel\propto\over\sim$}}\ }

\begin{document}

\title{Interaction in the dark sector: a Bayesian analysis with latest observations}

\author{T. Ferreira$^{1}$\email{tassia.aferreira@gmail.com}, C. Pigozzo$^{2}$\email{kssiobr@gmail.com }, S. Carneiro$^{1,2}$\email{saulo.carneiro.ufba@gmail.com }, J. S. Alcaniz$^{3,4}$\email{alcaniz@on.br}}

\affiliation{$^1$PPGCosmo, CCE, Universidade Federal do Esp\'irito Santo, 29075-910, Vit\'oria, ES, Brasil}
\affiliation{$^2$Instituto de F\'{\i}sica, Universidade Federal da Bahia, 40210-340, Salvador, BA, Brasil}
\affiliation{$^3$Observat\'orio Nacional, 20921-400, Rio de Janeiro, RJ, Brasil}
\affiliation{$^4$Physics Department, McGill University, Montreal QC, H3A 2T8, Canada}

\date{\today}

\begin{abstract}
By combining cosmological probes at low, intermediate and high redshifts, we investigate the observational viability of a class of models with interaction in the dark sector. We perform a Bayesian analysis using the latest data sets of type Ia supernovae, baryon acoustic oscillations, the angular acoustic scale of the cosmic microwave background, and measurements of the expansion rate. When combined with the current measurement of the local expansion rate obtained by the Hubble Space Telescope, we find that these observations provide evidence in favour of interacting models with respect to the standard cosmology. 
\end{abstract}

\pacs{98.65.Dx, 98.80.Es}
\maketitle


\section{Introduction}

The Universe we observe today has been largely explained by the $\Lambda$ plus cold dark matter ($\Lambda$CDM) cosmology.  There are, however, some important theoretical issues in the context of this scenario that need to be better understood as, for instance, the value of the cosmological constant $\Lambda$, whose theoretical expectations differ from the observational value up to $~{120}$ orders of magnitude~\cite{Weinberg:1988cp, Sahni:1999gb, Padmanabhan:2002ji}. From the observational side, there are also some tensions which emerge when different data sets at different scales are analysed. Examples are the current  value of the Hubble parameter, $H_0$, estimates of the power spectrum amplitude on scales of $8h^{-1}$ Mpc, $\sigma_8$, and measurements of the matter density parameter, $\Omega_m$, (see, e.g., \cite{Macaulay:2013swa, Freedman:2017yms, Amon:2017lia, Benetti:2017juy} and the references therein for details). 

Some of these issues have motivated the investigation of alternative models of the Universe, usually associated with physical processes involving either new fields in high energy physics or modifications of gravity on large scales~\cite{Copeland:2006wr, Capozziello:2002rd, Alcaniz:2006ay, Frieman:2008zz}. In principle, to check the validity of a theory or model (for instance, the standard $\Lambda$CDM cosmology), it is always interesting to insert it in a more general framework. Here we study a class of phenomenological models named non-adiabatic generalised Chaplygin gas (GCG), of which the standard model is a particular case \citep{zimdahl,gcg:carneiro14,pigozzo16}. Due to their generality, these models  deserve a broader and careful discussion. As can be shown, the $\Lambda$CDM results are readily recovered for a specific value of the dimensionless parameter $\alpha$  ($\alpha = 0$). Another well-motivated class of models in which dark energy decays into dark matter at a constant rate (hereafter denoted by $\Lambda(t)$CDM) is fully described for $\alpha = -1/2$~\cite{gcg:borges05,PLB2012}.

In what follows, we provide a brief revision of the GCG model, highlighting its dynamics and evolution, and also showing how it can be reduced to either the $\Lambda$CDM and $\Lambda(t)$CDM models. Next, we introduce the observational data which are used to perform a Bayesian analysis of these models, namely: $740$ type Ia supernovae (SNe Ia) from the Joint Light-curve Analysis (JLA) dataset~\cite{sne:betoule14}, $4$ measurements of baryon acoustic oscillations (BAO) using the ratio $D_V/r_s$, obtained from two-point correlation function (2PCL) analysis~\cite{bao:beutler11, bao:blake11}, $14$ measurements of BAO ($\theta_{BAO}$), obtained using an angular two-point correlation function (2PACL)~\cite{carvalho16, alcaniz17, carvalho17, ecarvalho17}, the acoustic scale $\ell_{\mathrm{A}}$ derived from the {Planck} 2015 anisotropy spectrum of the cosmic microwave background (CMB) \cite{planck15}, and  $25$ data points of the expansion rate, $H(z)$, obtained from red galaxies up to $z \sim 1$~\cite{simon05, stern10, moresco12, moresco16}. The parameter space is explored by using the M{\small ULTI}N{\small EST} algorithm~\cite{nested:feroz09,nested:feroz13}. It also provides the Bayesian evidence as a by-product, which is used to compute the Bayes' factor and indicates which model best reproduces the observations. Our analysis shows that the current cosmological data provide evidence in favour of interacting models with respect to the standard $\Lambda$CDM cosmology. Throughout this paper we work with units where $c = 8 \pi G = 1$.


\section{Cosmological Models}
\thispagestyle{empty}

In this section we briefly present the cosmological models used in our analysis. We will introduce the non-adiabatic  generalised Chaplygin gas model and then the two of its particular cases mentioned earlier, i.e.,  the $\Lambda$CDM and $\Lambda(t)$CDM models.

\subsection{Non-adiabatic generalised Chaplygin gas}

Let us start by considering that the dark sector of the Universe is composed of a fluid with negative pressure described by~\cite{Kamenshchik:2001cp,fabris,fabris2, Bilic:2001cg, Bento:2002ps, Dev:2002qa, Alcaniz:2002yt}
\begin{equation} \label{dark-pressure}
p = - \frac{A}{\rho^{\alpha}} ,
\end{equation}
where $A$ is a positive constant and $\alpha > -1$. Its equation of state is $p = \omega \rho$, with $-1 < \omega < 0$. This dark fluid can be formally broken into two components: a non-relativistic, pressureless fluid, which can cluster, known as dark matter; and a dark energy component with $\omega_{\Lambda} = -1$, such that we can write $p_{\Lambda} = - \rho_{\Lambda}$ \citep{carneiro14}.
For a spatially flat Universe, the Friedmann and energy balance equations can be written, respectively, as
\begin{equation} \label{dark-fried}
\rho = \rho_{dm} + \rho_{\Lambda} = 3 H^2\; ,
\end{equation}
and
\begin{equation}
\label{dark-eoc}
\dot{\rho}_{dm} + 3 H \rho_{dm} = - \dot{\rho}_{\Lambda} \;.
\end{equation}
Notice that Eq. (\ref{dark-eoc}) describes how the two components of the dark sector interact. We can use Eq. \ref{dark-fried} to show that
\begin{equation} \label{rho-lambda}
\rho_{\Lambda} = -3 \omega H^2\; .
\end{equation}
Also, if we differentiate Eq. (\ref{dark-fried}) and substitute it in Eq. (\ref{dark-eoc}) we find
\begin{equation} \label{rho-m}
2 \dot{H} = - \rho_{dm} \;.
\end{equation}
Now we can combine Eq. (\ref{rho-lambda}) with Eq. (\ref{rho-m}) to write
\begin{equation} \label{gamma}
\Gamma = \frac{\dot{\omega}}{1 + \omega } - 3 \omega H\;,
\end{equation}
where we introduce $\Gamma = - \dot{\rho}_{\Lambda} / \rho_{dm} $ as the energy flux rate, used to define the rate at which dark matter is created.
The adiabatic speed of sound for the generalised Chaplygin gas is given by \citep{wands12,wang13}
\begin{equation}
c^{2}_{a} = \frac{\dot{p}}{\dot{\rho}} = \alpha \omega\;,
\end{equation}
from where we obtain 
\begin{equation}
\dot{\omega} \rho = \omega(\alpha + 1)\dot{\rho}\; .
\end{equation}
If we use this in Eq. (\ref{dark-eoc}) we find
\begin{equation}
\dot{\omega} = 3 \omega (\alpha + 1)(\omega + 1)H\; .
\end{equation}
This result can be substituted in our definition of the energy-flux rate to rewrite it as
\begin{equation}
\Gamma = 3 \alpha \omega H \;.
\end{equation}
With the help of Eq. \ref{dark-pressure}, we find
\begin{equation}
\omega = \frac{p}{\rho} = - \frac{A}{\rho^{\alpha + 1}} \;.
\end{equation}
Hence, if we recall Eq. \ref{dark-fried}, we have
\begin{equation}
\Gamma = - \frac{\alpha A}{3^{\alpha}}\  H^{-(2 \alpha + 1)} \;.
\end{equation}
We can see that the value of $\alpha$ characterises the interaction between the two components of the dark sector. For $\alpha < 0$, we have $\Gamma > 0$, which implies that there is an energy flux from dark energy to dark matter. If $\alpha = 0$, we are reduced to the $\Lambda$CDM model, where $\Gamma = 0$. On the other hand, for $\alpha = - 1/2$, $\Gamma$ is constant and we obtain the $\Lambda(t)$CDM model, as we will see in Section \ref{ltcdm}.

Let us now turn to the function which describes the dynamics and evolution of this model, $E(z) = H(z)/H_{0}$, where a subscript `0' denotes present-day quantities. If we substitute Eq. (\ref{dark-pressure}) in Eq. (\ref{dark-eoc}), it gives
\begin{equation}
\frac{\rho^{\alpha}}{\rho^{\alpha +1} - A}\ \dot{\rho} = - 3 \frac{\dot{a}}{a}\ ,
\end{equation}
and integrating it leads to
\begin{equation}
\rho = \rho_0 \bigg[ (1 - \bar{A}) \bigg( \frac{a_0}{a} \bigg)^{3(1+ \alpha)} + \bar{A} \bigg]^{\frac{1}{1+ \alpha}} \;,
\end{equation}
where $\bar{A} \equiv A/\rho_{0}^{\alpha + 1}$. If we replace $\rho = 3 H^2$ in the above equation and use $a_0 = 1$ it becomes
\begin{equation}
E(z) = \big[ (1 - \bar{A})(1 + z)^{3(1+ \alpha)} + \bar{A} \big]^{\frac{1}{2(1+ \alpha)}} \;.
\end{equation}
Taking the limit $z \rightarrow -1$, we notice that $\bar{A} = \Omega_{\Lambda,0} = 1 - \Omega_{m,0}$. We have also $\Omega_{\Lambda,0} = - \omega_0$, since
\begin{equation}
\omega_0 = \frac{p_{m,0} + p_{\Lambda,0}}{\rho_{m,0} + \rho_{\Lambda,0}} = - \frac{\rho_{\Lambda,0}}{\rho_{c,0}} ,
\end{equation}
with $\rho_{c,0}$ being the value of the critical density today, and, by definition, $\rho_{\Lambda,0}/\rho_{c,0} = \Omega_{\Lambda,0}$.
To include the contribution of the radiation component, $\Omega_{r,0}$, we simply add a term proportional to $(1+z)^{-4}$, since it is conserved separately. Finally, we have
\begin{equation} \label{eq-darkness}
E(z) = \sqrt{\big[ \Omega_{m,0}(1 + z)^{3(1+ \alpha)} + \Omega_{\Lambda,0} \big]^{\frac{1}{1+ \alpha}} + \Omega_{r,0}(1+z)^4}\ .
\end{equation}
Through a binomial expansion of this function, we find a term scaling as 
\begin{equation}
3 H_{0}^2 \Omega_{m,0}^{\frac{1}{1+ \alpha}}(1 + z)^{3}\;,
\end{equation}
which gives the matter density at high redshifts. This assures us that the term $\Omega_{m,0}$ in Eq. (\ref{eq-darkness}) also includes the conserved baryons.

\subsection{$\Lambda$CDM}

In the $\Lambda$CDM scenario, the two dark components are separately conserved. If we consider a spatially flat Universe, we have
\begin{equation}
E(z) = \sqrt{\Omega_{m,0}(1 + z)^{3} +  \Omega_{\Lambda,0} + \Omega_{r,0}(1+z)^4}\;,
\end{equation}
which is equivalent to setting $\alpha = 0$ in Eq. (\ref{eq-darkness}).

\subsection{$\Lambda(t)$CDM} \label{ltcdm}

The $\Lambda(t)$CDM model proposed in \cite{gcg:borges05,PLB2012} attempts to alleviate the tension between the value of dark energy we observe today and that obtained in quantum field theories. This is done by taking into account the contributions of interacting fields in the low-energy limit, which gives us a dark energy density scaling as \citep{schutz02},
\begin{equation}
\rho_{\Lambda} = m^3 H ,
\end{equation}
where $m \equiv 150$ MeV is the energy scale of the chiral phase transition of QCD. 
The conservation equation for this model is similar to that of the GCG, where dark energy decays into dark matter, as given by Eq. \ref{dark-eoc}. We must postulate that only dark matter is produced and that baryonic matter is conserved, as not to have problems with primordial nucleosynthesis. The Hubble parameter evolution for this model is given by
\begin{equation}
E(z) = \sqrt{ \big [\Omega_{m,0}(1 + z)^{3/2} + \Omega_{\Lambda,0} \big]^2 + \Omega_{r,0}(1 + z)^4}\,
\end{equation}
which is equivalent to taking $\alpha = -1/2$ in Eq. (\ref{eq-darkness}).
It is interesting to note that it has the same number of parameters as the standard model, which makes comparing them all the easier, since models with more parameters tend to be penalised in Bayesian statistics.


\section{Observational Probes}
\thispagestyle{empty}

When selecting cosmological probes, we must be careful to choose those in which we are able to separate their intrinsic evolution from the way they evolve with the redshift. For this purpose, we are mostly interested in probes which can be standardised, such as SNe Ia and  BAO measurements. Another factor to consider is how dependent a particular observation is on the fiducial cosmological model chosen to calibrate the data. This is one of our main concerns, since the purpose of this paper is to test alternatives to the standard model. Out of the four probes we have used here, the cosmological clocks currently present the largest source of systematic errors. For this reason, our joint analyses are made both with and without them, so that we may try to investigate whether they bias the result to a particular model.

\subsection{Type Ia Supernova}

Type Ia supernovae are one of the most important tools for studying the cosmic evolution of the Universe, and hence for constraining cosmological parameters. In our analysis, we have employed the JLA dataset, with 740 SNe Ia with redshifts up to $z \sim 1.3$ \citep{sne:betoule14}, and followed the procedure given in \cite{guy10} for obtaining the apparent magnitude in the B-band:
\begin{equation}
m_{B}^{\mathrm{mod}} = 5 \log_{10} d_L(z_{\mathrm{CMB}}, z_{\mathrm{hel}}) - \alpha X_1 + \beta \mathcal{C} + \mathcal{M}_B ,
\end{equation}
where $z_{\mathrm{CMB}}$ and $z_{\mathrm{hel}}$ are the redshift in the CMB rest frame and the heliocentric redshift, respectively. The parameters that model the light-curves are $X_1$ and $\mathcal{C}$, which are responsible for characterising the stretch and variation in colour, while $\alpha$ and $\beta$ are nuisance parameters. $\mathcal{M}_B$ is also a nuisance parameter, corresponding to the magnitude of a fiducial supernova, and $d_L$ is the luminosity distance.

We construct the $\chi^2$ as
\begin{equation}
\chi_{SN}^{2} = \sum (\mathrm{\mathbf{m_{B}}}-\mathrm{\mathbf{m_{B}^{mod}}})^{\mathrm{\mathbf{T}}} (\mathrm{\mathbf{C^{-1}_{SN}}})(\mathrm{\mathbf{m_{B}}}-\mathrm{\mathbf{m_{B}^{mod}}}) ,
\end{equation}
where the covariance matrix is given by
\begin{equation}
\mathrm{\mathbf{C}_{\mathbf{SN}}} = \mathrm{\mathbf{D}_{stat}} + \mathrm{\mathbf{C}_{stat}} + \mathrm{\mathbf{C}_{sys}} ,
\end{equation}
with
\begin{equation}
\begin{split}
\mathrm{\mathbf{D}_{stat, ii}} & = \bigg[ \frac{5}{z_i \ln 10}\bigg]^2 \sigma^{2}_{z,i} + \sigma^{2}_{lens} + \sigma^{2}_{int} \\
& + \sigma^{2}_{m_{B},i}  + \alpha^2 \sigma^{2}_{X_{1},i} + \beta^2 \sigma^{2}_{\mathcal{C},i} \\
& + 2 \alpha C_{m_{B}X_{1},i} - 2 \beta C_{m_{B} \mathcal{C}, i} - 2 \alpha \beta C_{X_{1} \mathcal{C},i} .
\end{split}
\end{equation}
The first three terms account for the errors in redshift due to peculiar velocities, the variation in the magnitude caused by lensing, and all other factors not already accounted by the former two, respectively.
The fourth term, $\sigma_{m_B}$, is the error of the observed magnitude. The other terms are errors in stretch ($X_1$), colour ($\mathcal{C}$), and their covariance matrices. The sum of the statistical and systematic covariance matrices is
\begin{eqnarray}
\mathrm{\mathbf{C}_{stat}} + \mathrm{\mathbf{C}_{sys}} & = & V_0 + \alpha^2 V_a + \beta^2 V_b + 2 \alpha V_{0a} \nonumber \\ & & - 2 \beta V_{0b} - 2 \alpha \beta V_{ab} ,
\end{eqnarray}
where $V_{0}, V_{a}, V_{b}, V_{0a}, V_{0b}$ and $V_{ab}$ are matrices provided by the sample~\cite{sne:betoule14}.

\subsection{Baryon Acoustic Oscillations}

\subsubsection{Measurements of $D_V/r_s$}

The BAO's  left mensurable signatures in the distribution of galaxies which have been robustly detected from data of galaxy redshift surveys~\cite{Aubourg:2014yra}.
From the currently available observations, angle-averaged clustering
data determine the ratio $D_V/r_s$, where $r_s$ is the acoustic horizon at the end of
radiation drag and $D_V$ is given by~\cite{Eisenstein:2005su}
\begin{equation}
D_V = \frac{c}{H_0} \bigg\{ \frac{z}{E(z)}\bigg[ \int_{0}^{z} \frac{dz'}{E(z')} \bigg]^2\bigg\}^{1/3}.
\end{equation}
Measurements of this type assume a fiducial cosmology in order to
transform the measured angular positions and redshifts
into comoving distances, which may bias the parameter
constraints~\cite{Carnero:2011pu, Salazar-Albornoz:2016psd}.
The data used here are given in Table \ref{tab:bao_cov}.
In order to constrain the cosmological parameters, we evaluate
\begin{equation}
\chi_{BAO}^{2} = \sum (\mathrm{\mathbf{x}}-\mathrm{\mathbf{d}})^{\mathrm{\mathbf{T}}} (\mathrm{\mathbf{C^{-1}_{BAO}}})(\mathrm{\mathbf{x}}-\mathrm{\mathbf{d}}) ,
\end{equation}
where $\mathrm{x_i}-\mathrm{d_i} = D_V(z_i)/r_s - d_{z,i}$, and $\mathrm{\mathbf{C^{-1}_{BAO}}}$ is the inverse of the covariance matrix.

	\begin{table}
	\centering
	\begin{tabular}{c c c}
	\hline
	\hline
	$z$   	& $d_z = D_V/r_s$            & References 			\\ \hline \\
	$0.106$ 	& $2.976 \pm 0.133$ 	& \cite{bao:beutler11}	\\[1ex]
	$0.440$ 	& $11.551\pm 0.559$ 	& \cite{bao:blake11}		\\[1ex]
	$0.600$ 	& $14.944\pm 0.677$ 	& \cite{bao:blake11}		\\[1ex]
	$0.730$ 	& $16.932\pm 0.580$ 	& \cite{bao:blake11}		\\[2ex]
	\hline
	\hline
	\end{tabular}
	\caption{The first column shows the redshift in which $D_V/r_s$ was measured.  The second column contains the mean values and the standard deviations of $D_V/r_s$. The third column shows the due references.}
	\label{tab:bao_cov}
	\end{table}

\subsubsection{Measurements of $\theta_{BAO}$}

A more model-independent way to measure the signature of the BAO's can be obtained from a 2-point angular correlation function analysis. The procedure consists of measuring the angular separation between pairs for a defined comoving acoustic scale, considering very thin redshift shells of order $0.01 - 0.02$~\citep{carvalho16}. Since there is no need for a fiducial cosmology, the measurements of  $\theta_{\mathrm{BAO}}$ are almost  model-independent, which makes this a robust tool for testing cosmological models\footnote{As discussed in Refs.~\cite{carvalho16}, in order to account for the projection effects introduced by the shell width, a theoretical model is assumed in the analysis, which introduces
a 1-2\% model-dependent correction in the $\theta_{BAO}$ position.}.
The theoretical value of $\Theta_{\mathrm{BAO}}$ for a given cosmological model can be found by employing
\begin{equation} \label{eq.bao}
\theta_{\mathrm{BAO, th}} =  \frac{r_s}{D_A}\ ,
\end{equation}
with
\begin{equation}
D_A = {H_0}^{-1} \int_{0}^{z} \frac{dz'}{E(z')} .
\end{equation}
For a model with a set of free parameters $\mathbf{S}$, we have
\begin{equation}
\chi_{\theta_{\mathrm{BAO}}}^{2} = \sum_{i} \frac{ \big[\theta_{\mathrm{BAO,obs}}(z_i) - \theta_{\mathrm{BAO},th}(z_i,\mathbf{S}) \big]^2}{\sigma^2_i},
\end{equation}
where $\theta_{\mathrm{BAO,obs}}$ are the values of $\theta_{\mathrm{BAO}}$ given in Table \ref{tab:theta}, $\theta_{\mathrm{BAO},th}$ is calculated using Eq. (\ref{eq.bao}), and $\sigma$ is the error, also given in the aforementioned table.

	\begin{table}
	\centering
	\begin{tabular}{c c c}
	\hline
	\hline
	$z$ 		& $\Theta_{\mathrm{BAO}} (^{\circ})$	& References 			\\ \hline \\
	$ 0.235 $ & $ 9.105 \pm 0.230$ 				& \cite{alcaniz17}		\\ [1ex]
	$ 0.365 $ & $ 6.362 \pm 0.220$	 				& \cite{alcaniz17}		\\ [1ex]
	$ 0.450 $ & $ 4.767 \pm 0.170$ 				& \cite{carvalho16}		\\ [1ex]
	$ 0.470 $ & $ 5.017 \pm 0.250$ 				& \cite{carvalho16}		\\ [1ex]
	$ 0.490 $ & $ 4.989 \pm 0.210$ 				& \cite{carvalho16}		\\ [1ex]
	$ 0.510 $ & $ 4.814 \pm 0.170$ 				& \cite{carvalho16}		\\ [1ex]
	$ 0.530 $ & $ 4.291 \pm 0.300$ 				& \cite{carvalho16}		\\ [1ex]
	$ 0.550 $ & $ 4.250 \pm 0.250$ 				& \cite{carvalho16}		\\ [1ex]
	$ 0.570 $ & $ 4.593 \pm 0.355$ 				& \cite{carvalho17}		\\ [1ex]
	$ 0.590 $ & $ 4.394 \pm 0.330$ 				& \cite{carvalho17}		\\ [1ex]
	$ 0.610 $ & $ 3.856 \pm 0.305$ 				& \cite{carvalho17}		\\ [1ex]
	$ 0.630 $ & $ 3.912 \pm 0.430$ 				& \cite{carvalho17}		\\ [1ex]
	$ 0.650 $ & $ 3.550 \pm 0.160$ 				& \cite{carvalho17}		\\ [1ex]
	$ 2.225 $ & $ 1.850 \pm 0.330$ 				& \cite{ecarvalho17}		\\ [2ex]
	\hline
	\hline
	\end{tabular}
	\caption{Values of angular baryon acoustic observations obtained by the cited authors, and their respective redshifts, used in this paper.}
	\label{tab:theta}
	\end{table}

\subsection{CMB Acoustic Scale}

The full analysis of the {Planck} 2015 provides a nearly model independent, well constrained value of the angular scale of the sound horizon at the time of decoupling, $\ell_{\mathrm{A}}$. Although it provides only one data point, it is useful when combined with other cosmological probes, since it carries important information about the early-universe physics. We have
\begin{equation}
\ell_{\mathrm{A}} = \frac{\pi}{\Theta_A} ,
\end{equation}
with $\Theta_A$ being the angular scale of the first peak of the angular power spectrum of CMB anisotropies, given by
\begin{equation}
\Theta_A = \bigg[ \int^{\infty}_{z_{*}} \frac{c_s (z)}{c} \frac{dz}{H(z)} \bigg] \bigg[ \int^{z_{*}}_{0} \frac{dz}{H(z)} \bigg]^{-1} .
\end{equation}
The limit $z_{*}$ refers to the redshift at the time of the last scattering. The ratio between the speed of sound $c_s$ and the speed of light can be calculated using
\begin{equation}
\frac{c_s (z)}{c} = \bigg[ 3 + \frac{9 \rho_b (z)}{4 \rho_\gamma (z)} \bigg]^{-1/2} ,
\end{equation}
where $\rho_b$ is the baryon density and $\rho_\gamma$ is the photon density. We get the redshift $z_{*}$ by computing \citep{hu96}
\begin{equation}
z_{*} = 1048 \times \big [ 1 + 0.00124 (\Omega_b h^2)^{-0.738} \big ] \big [ 1 + g_1(\Omega_m h^2)^{g_2} \big ] ,
\end{equation}
\begin{equation}
g1 = \frac{0.0783 (\Omega_b h^2)^{-0.238}}{1 + 39.5 (\Omega_b h^2)^{0.763}} ,
\end{equation}
\begin{equation}
g2 = \frac{0.560}{1 + 21.1 (\Omega_b h^2)^{1.81}} ,
\end{equation}
where $\Omega_b$ and $\Omega_m$ are the relative densities of baryons and total matter, respectively.
The value of $\ell_{\mathrm{A}} =  301.63 \pm {0.15}$ used in this analysis is taken from {Planck} 2015 results \citep{planck15}. 

\subsection{Cosmological Clocks}

Differently from the other probes discussed earlier cosmological clocks are based upon measurements of time over the cosmic evolution. The Observational Hubble Data (OHD) is obtained by using the differential age method \citep{jimenez02, dunlop96}, which constitutes of taking the relative age of passively evolving galaxies $\Delta t$, separated by a redshift $\Delta z$. This is then used to obtain
\begin{equation}
H(z) = - \frac{1}{1+z} \frac{dz}{dt} .
\end{equation}
However simple this equation may be, obtaining the ratio $\Delta z/ \Delta t$ is rather difficult, because one needs a good sample of populations of galaxies, and also has to rely largely on stellar population synthesis models~\citep{moresco12}. Our dataset consists of values calculated by several authors, as shown in Table \ref{tab:h_clocks}.
The $\chi_{OHD}^{2}$ is given by
\begin{equation}
\chi_{OHD}^{2} = \sum_{i} \frac{\big[\mathrm{OHD}_{\mathrm{obs}}(z_i) - \mathrm{OHD}_{\mathrm{th}}(z_i,\mathbf{S}) \big]^2}{\sigma^2_i} ,
\end{equation}
where $\mathrm{OHD}_{\mathrm{obs}}$ is the value of OHD obtained from the sample, with its respective error $\sigma_i$, and $\mathrm{OHD}_{\mathrm{th}}$ is the calculated value for each model. We add to this $H(z)$ sample the current value of the Hubble parameter, $H_0 = 73.24 \pm 1.74$ km/s/Mpc, derived by Riess {\it{et al.}}~\cite{riess16} using four geometric distance calibrations of Cepheids.

	\begin{table}
	\centering
	\begin{tabular}{c c c}
	\hline
	\hline
	$z$     	& OHD  			& References 		\\ \hline \\
	$0$		& $ 73.24 \pm 1.74 $& \cite{riess16}		\\ [1ex]
	$0.07$ 	& $ 69    \pm 19.6 $	& \cite{moresco16}	\\ [1ex]
	$0.09$ 	& $ 69    \pm 12   $	& \cite{moresco16}	\\ [1ex]
	$0.12$ 	& $ 68.6  \pm 26.2 $	& \cite{moresco16}	\\ [1ex]
	$0.17$ 	& $ 83    \pm 8    $	& \cite{simon05}	\\ [1ex]
	$0.179$	& $ 75    \pm 4    $	& \cite{moresco12}	\\ [1ex]
	$0.199$ 	& $ 75    \pm 5    $	& \cite{moresco12}	\\ [1ex]
	$0.20$ 	& $ 72.9  \pm 29.6 $	& \cite{zhang14}	\\ [1ex]
	$0.27$ 	& $ 77    \pm 14   $	& \cite{simon05}	\\ [1ex]
	$0.28$ 	& $ 88.8  \pm 36.6 $	& \cite{zhang14}	\\ [1ex]
	$0.352$ 	& $ 83    \pm 14   $	& \cite{moresco12}	\\ [1ex]
	$0.3802$ 	& $ 83    \pm 13.5 $	& \cite{moresco16}	\\ [1ex]
	$0.4$ 	& $ 95    \pm 17   $	& \cite{simon05}	\\ [1ex]
	$0.4004$ 	& $ 77    \pm 10.2 $	& \cite{moresco16}	\\ [1ex]
	$0.4247$ 	& $ 87.1  \pm 11.2 $	& \cite{moresco16}	\\ [1ex]
	$0.44497$& $ 92.8  \pm 12.9 $	& \cite{moresco16}	\\ [1ex]
	$0.4783$ 	& $ 80.9  \pm 9    $	& \cite{moresco16}	\\ [1ex]
	$0.48$ 	& $ 97    \pm 62   $	& \cite{stern10}		\\ [1ex]
	$0.593$ 	& $ 104   \pm 13   $	& \cite{moresco12}	\\ [1ex]
	$0.68$ 	& $ 92    \pm 8    $	& \cite{moresco12}	\\ [1ex]
	$0.781$ 	& $ 105   \pm 12   $	& \cite{moresco12}	\\ [1ex]
	$0.875$ 	& $ 125   \pm 17   $	& \cite{moresco12}	\\ [1ex]
	$0.88$ 	& $ 90    \pm 40   $	& \cite{stern10}		\\ [1ex]
	$0.9$ 	& $ 117   \pm 23   $	& \cite{simon05}	\\ [1ex]
	$1.037$ 	& $ 154   \pm 20   $	& \cite{moresco12} 	\\ [2ex]
	\hline
	\hline
	\end{tabular}
	\caption{Values of the observational Hubble data obtained by different authors using the differential age method. The first point corresponds to the locally measured value of the Hubble constant.}
	\label{tab:h_clocks}
	\end{table}


\section{Statistical Analysis}

Bayesian inference has become one of the most popular methods for analysing data in cosmology (see, e.g.,~\cite{Heavens:2017hkr, SantosdaCosta:2017ctv, Andrade:2017iam} and references therein). It provides a way to easily deal with the problem of nuisance parameters by giving us the freedom to marginalise over them. It also takes into account prior information, which can be very useful in the case of parameters that have a physical meaning, or when we have already obtained their value by using a model-independent approach. 
For these reasons, we have chosen to use Bayesian inference while employing the M{\small ULTI}N{\small EST} algorithm \citep{nested:feroz09,nested:feroz13}, which is simple to use and can be modified to work with different cosmological models and various observational probes. To ensure that M{\small ULTI}N{\small EST} was free of bias to any particular model, we have previously carried out the tests with our own codes, and only then switched over to gain computational time and accuracy. The priors we have used are detailed in Table \ref{tab:priors}. For all other parameters not mentioned in the table, we adopted a  flat prior.

	\begin{table}
	\centering
	\begin{tabular}{c c c}
	\hline
	\hline
	Parameter			&	Prior				& References	 	\\ \hline \\
	$\Omega_bh^2$	& $0.02226 \pm 0.00023$	& \cite{planck15}	\\ [1ex]
	$r_s$			& $141.1 \pm 5.5$ {Mpc}		&  \cite{licia17}		\\ [1ex]
	$h$				& $0.7324 \pm 0.0174$	&  \cite{riess16}		\\ [2ex]
	\hline
	\hline
	\end{tabular}
	\caption{Gaussian priors used in the analysis.}
	\label{tab:priors}
	\end{table}

\subsection{Bayesian Evidence}

When comparing models, the Bayesian evidence is crucial since it holds information on which of the models are best able to reproduce the observations. Again,  M{\small ULTI}N{\small EST} proved to be the best choice for this analysis, because it gives us the evidence as a by-product of its calculations. 
 For a given set of data $\mathbf{D}$, described by a model $M$, with parameters $\mathbf{S}$, we have Bayes' Theorem
\begin{equation}
\mathcal{P}(\mathbf{S} | \mathbf{D}, M) = \frac{\mathcal{P}(\mathbf{D} | \mathbf{S}, M)\ \mathcal{P}(\mathbf{S} | M) }{ \mathcal{P}(\mathbf{D} | M)},
\end{equation}
where the posterior probability, $\mathcal{P}(\mathbf{S} | \mathbf{D}, M)$, holds all we know about $\mathbf{S}$, after we have analysed $\mathbf{D}$; the likelihood, $\mathcal{P}(\mathbf{D} | \mathbf{S}, M)$, tells us the probability of reproducing the data for different values of the parameters; the prior is represented by $\mathcal{P}(\mathbf{S} | M)$; and, lastly, the denominator $\mathcal{P}(\mathbf{D} | M) \equiv \mathcal{Z}$ is the evidence.
Once we have the evidence, we are able to apply Bayes' Factor to determine which of the two models is more favoured by the data,
\begin{equation}
B(\mathbf{D}) = \frac{\mathcal{P}(\mathbf{D} | M_{1})}{\mathcal{P}(\mathbf{D} | M_{2})} .
\end{equation}
A value $B(\mathbf{D})>1$ favours $M_{1}$, otherwise $M_{2}$ is favoured. A further step involves invoking Jeffreys' scale, to determine how much confidence we can have in our results. 
We use the revisited version of the Jeffreys scale suggested in \cite{Trotta:2005ar}: ${\textit{inconclusive}}$ for $|\ln \mathit{B}|  = 0 - 1$,  ${\textit{weak}}$ for $| \ln \mathit{B}|  = 1 - 2.5 $,  ${\textit{moderate}}$  for $|\ln \mathit{B}|  = 2.5 - 5 $ and ${\textit{strong}}$ for $| \ln\mathit{B}|  > 5 $. Note that a negative $\ln \mathit{B}$ means  preference of the reference ($M_{1}$) over the analysed model ($M_{2}$). 


	\begin{table} [t]
	\centering
	\begin{tabular} { l c c} 
	\hline
	\hline
	Models			&		$h$					&			$\Omega_{m,0}$	\\ \hline \\
	GCG				&	$0.735^{+0.025}_{-0.024}$	&	$0.333^{+0.088}_{-0.099}$	\\ [1.5ex]
	$\Lambda(t)$CDM	&	$0.741^{+0.029}_{-0.029}$	&	$0.378^{+0.039}_{-0.036}$	\\ [1.5ex]
	$\Lambda$CDM	&	$0.728^{+0.027}_{-0.026}$	&	$0.243^{+0.041}_{-0.040}$	\\ [2ex]
	\hline
	\hline
	\end{tabular}
	\caption{Cosmological parameters obtained from the joint analysis of $H_0$+SN+BAO+$\Theta_{\mathrm{BAO}}+\ell_{\mathrm{A}}$ within a $2 \sigma$ interval.}
	\label{tab:parameters}
	\end{table}

	\begin{table}
	\centering
	\begin{tabular} { l c c c} 
	\hline
	\hline
	Data			&	$h$						&	$\Omega_{m0}$			&	$\alpha$				\\ \hline \\
	OHD+SN+BAO+$\ell_{\mathrm{A}}$	&	$0.722^{+0.022}_{-0.021}$	&	$0.292^{+0.086}_{-0.090}$	&	$-0.16^{+0.30}_{-0.29}$	\\ [1ex]
	$H_0$+SN+BAO+$\ell_{\mathrm{A}}$&	$0.735^{+0.025}_{-0.024}$	&	$0.333^{+0.088}_{-0.099}$	&	$-0.35^{+0.30}_{-0.26}$	\\ [2ex]
	\hline
	\hline
	\end{tabular}
	\caption{Cosmological parameters obtained from the joint analysis of OHD+SN+BAO+$\ell_{\mathrm{A}}$ and $H_0$+SN+BAO+$\ell_{\mathrm{A}}$ within an interval of $2 \sigma$.}
	\label{tab:constraints}
	\end{table}

\section{Results and discussion}

\begin{figure}[t]
	\centering
	\includegraphics[scale=0.32]{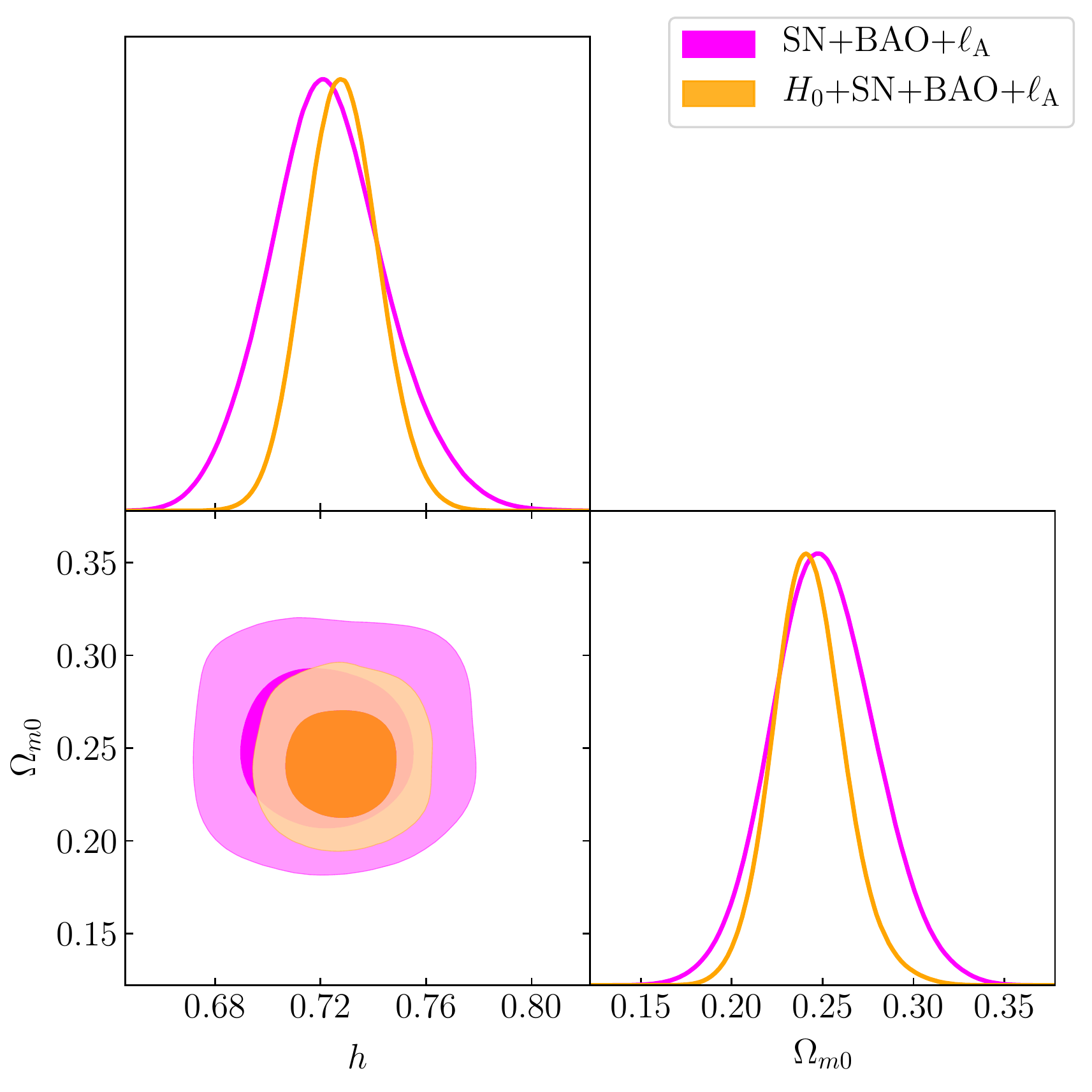} 
	\caption{The graph illustrates how the $H_0$ prior contributes to the constraining power over the models parameters. The PDF and contour plots are for the $\Lambda$CDM model, but the same behaviour holds for the two other models considered in the analysis.}
	\label{fig:clocks-comparison}
	\end{figure}

\begin{figure}[t]
	\centering
	\includegraphics[scale=0.32]{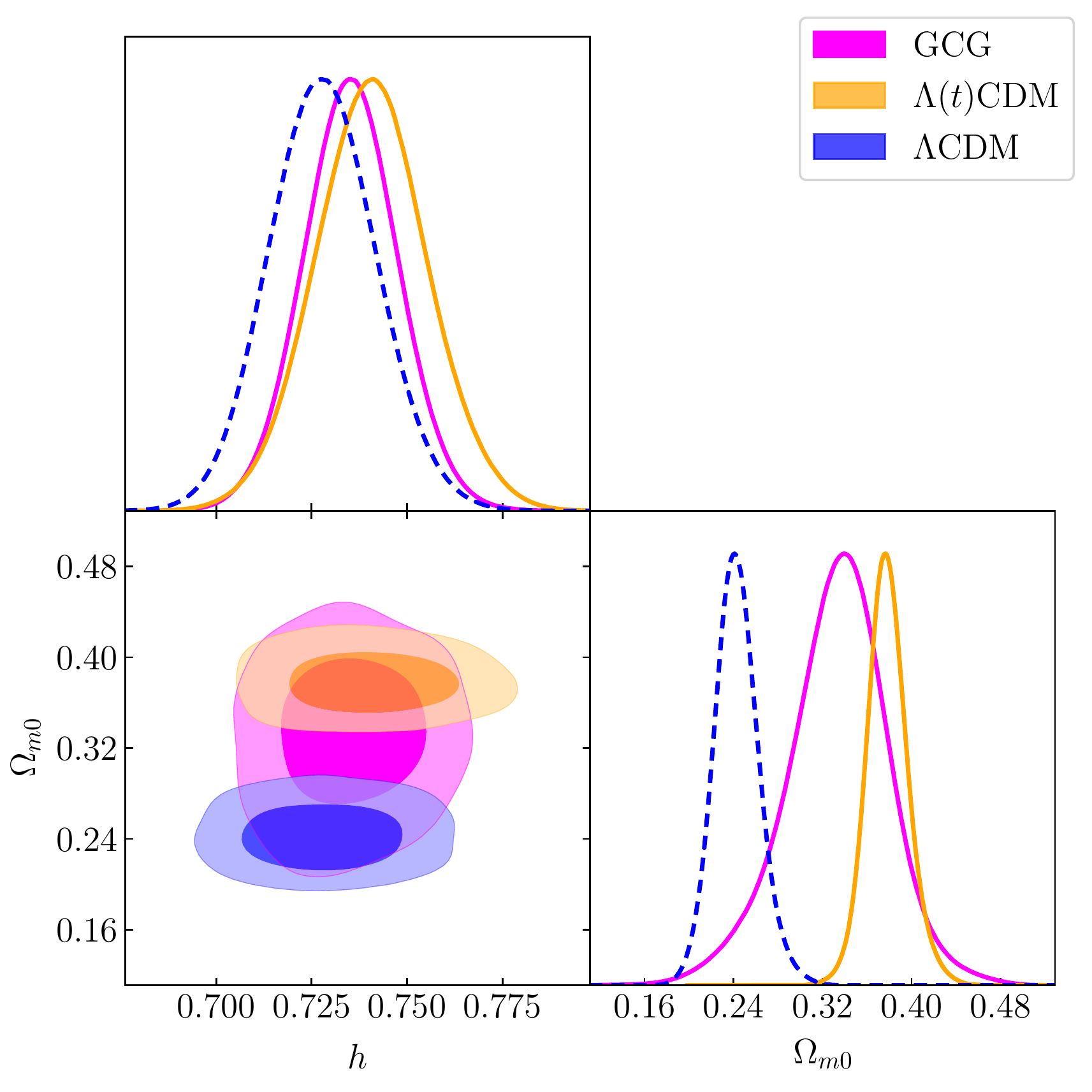}
	\caption{Contour plots and PDFs for the parameters $h$ and $\Omega_{m,0}$, for the GCG (fuchsia), $\Lambda(t)$CDM (orange) and $\Lambda$CDM (blue) models using tests with $H_0$+SN+BAO+$\ell_{\mathrm{A}}$.}
	\label{fig:h-wm}
	\end{figure}

	\begin{figure}[t!]
	\centering
	\includegraphics[scale=0.3]{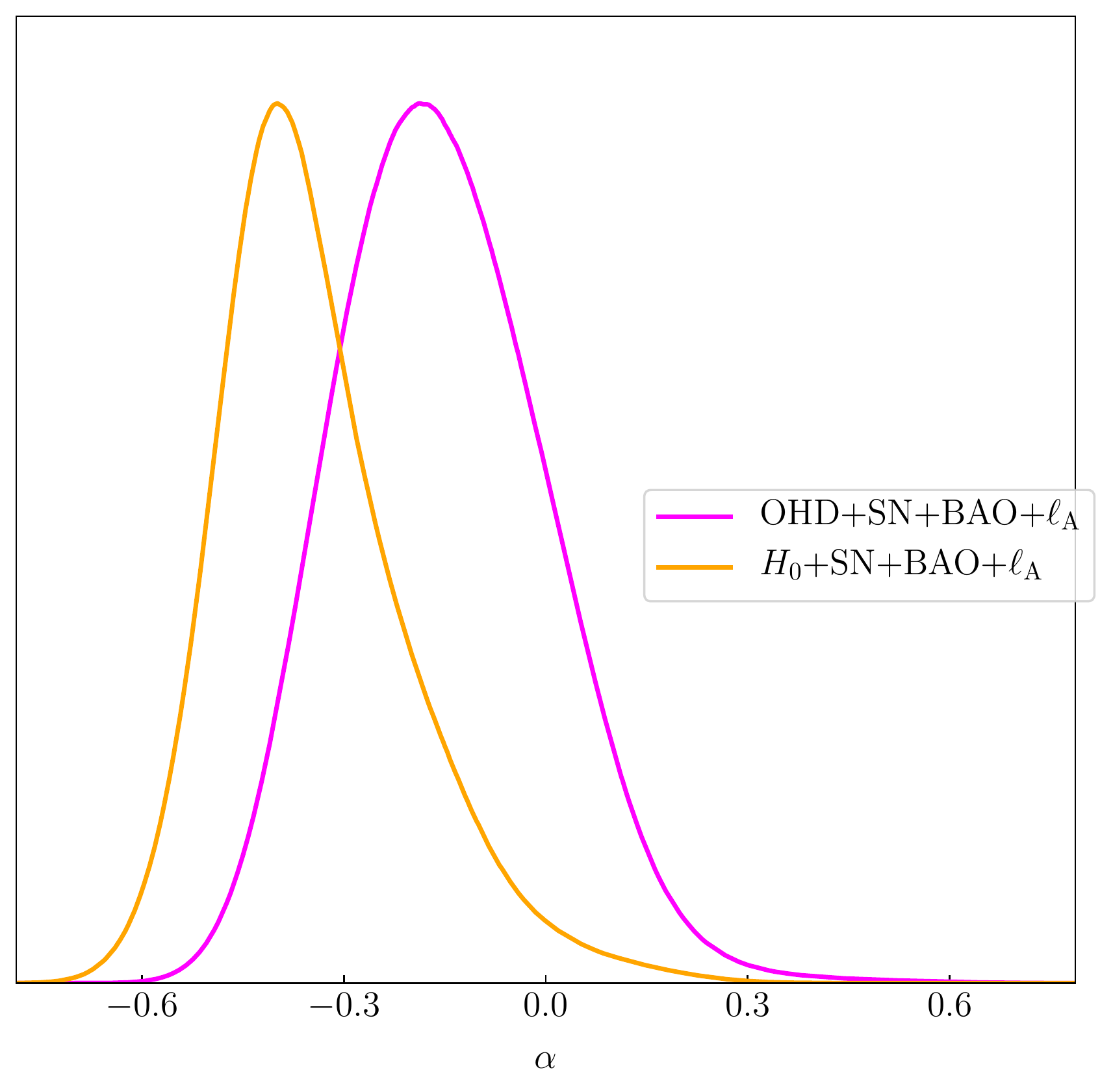}
	\caption{PDF of $\alpha$ for the non-adiabatic generalised Chaplygin gas model, for the joint analysis of OHD+SN+BAO+$\ell_{\mathrm{A}}$ (in fuchsia) and $H_0$+SN+BAO+$\ell_{\mathrm{A}}$ (in orange).}
	\label{fig:Chaplygin-alpha}
	\end{figure}
	
	
	\begin{table}
	\centering
	\begin{tabular} { l c c c} 
	\hline
	\hline
	Models			&		$h$					&			$\Omega_{m,0}$ 	&	$r_s$	\\ \hline \\
	GCG				&	$0.732^{+0.028}_{-0.025}$	&	$0.308^{+0.088}_{-0.089}$	  & $145.3^{+6.1}_{-6.1}$	\\ [1.5ex]
	$\Lambda(t)$CDM	&	$0.742^{+0.027}_{-0.030}$	&	$0.378^{+0.029}_{-0.027}$	  & $141.2^{+5.2}_{-4.8}$	\\ [1.5ex]
	$\Lambda$CDM	&	$0.723^{+0.026}_{-0.025}$	&	$0.249^{+0.028}_{-0.030}$	  & $148.8^{+4.6}_{-4.8}$	\\ [2ex]
	\hline
	\hline
	\end{tabular}
	\caption{Values of the cosmological parameters obtained from the joint analysis of $H_0$+SN+BAO+$\Theta_{\mathrm{BAO}}+\ell_{\mathrm{A}}$ within a $2 \sigma$ interval, when a flat prior is used for $r_s$.}
	\label{tab:flat}
	\end{table}

For tests using SNe Ia data, we have included the $H_0$ prior shown in Table \ref{tab:priors}. This procedure is advisable due to the degeneracies on $h$ and $\mathcal{M}_B$ when both are free parameters. Moreover, the best fit values are compatible with those when this prior is not used, as we can see in Figure \ref{fig:clocks-comparison} for the $\Lambda$CDM case. Without the prior, the best fit values are $h = 0.723^{+0.047}_{-0.042}$ and $\Omega_{m,0}	= 0.251^{+0.057}_{-0.054}$, whereas, when we employ it, we obtain $h = 0.728^{+0.027}_{-0.026}$ and $\Omega_{m,0} = 0.243^{+0.041}_{-0.040}$.

Table \ref{tab:parameters} displays the results for the analysis using $H_0$+SN+BAO+$\ell_{\mathrm{A}}$ (for simplicity, we will refer to the combined analysis of the two kinds of BAO measurements as BAO only), where we can see that the $\Lambda(t)$CDM model yields the highest value of $\Omega_{m,0}$, which is expected, since we are assuming a production of dark matter with the decay of dark energy. Similarly, for tests involving the GCG model, $\Omega_{m,0}$ is $37$\% greater than that obtained for the $\Lambda$CDM scenario. These results can also be seen in Figure \ref{fig:h-wm}, where we display the Probability Density Functions (PDF) and contour plots of $h$ and $\Omega_{m,0}$ for the three models.
When comparing the models, it is clear that those with interaction in the dark sector are more favoured by the observational probes. For the tests with $H_0$+SN+BAO+$\ell_{\mathrm{A}}$, there is a positive evidence ($1.627 \pm 0.518$) for the $\Lambda(t)$CDM model over the standard cosmology. On the other hand, when we look at $\Lambda$CDM and the GCG model, we have a moderate evidence ($3.314 \pm 0.513$) towards the latter. 

If we turn to the results obtained when we add the data from cosmological clocks, there is also a positive evidence ($1.362 \pm 0.513$) for the GCG model. However, in the case of $\Lambda(t)$CDM, this model is disfavoured against $\Lambda$CDM ($1.247 \pm 0.514$). The $2\sigma$ error interval, nevertheless, is large enough to encompass the inconclusive case as well, therefore it is safe to say that this particular dataset is incapable of  differentiating between the three models. 

Table \ref{tab:models} shows the complete analysis using the Bayes' Factor and Jeffreys' scale to compare the models to our selected set of combined observational probes. 
It is worth noting that cosmological clocks seem to shift the balance in favour of the standard model (lower values of $\Omega_{m,0}$, tendency towards a higher best fit value of $\alpha$, as shown in Table \ref{tab:constraints} and Figure \ref{fig:Chaplygin-alpha}, and the positive/inconclusive evidence towards $\Lambda$CDM), whereas, if we do not include this cosmological probe, the results do not disfavour the alternative models, but rather give conclusive evidence towards them. 
For tests carried out with each individual cosmological probe used here, most produced inconclusive results, except for cosmological clocks, which favoured the standard model, and the ones with only $\Theta_{\mathrm{BAO}}$, that gave a strong evidence ($\sim 6.97 \pm 0.122$) towards both models with interaction in the dark sector. 

Finally, it is worth mentioning that to obtain the above results we have used a model-independent Gaussian prior on the acoustic scale $r_s$ (see Table \ref{tab:priors}) derived from a subset of data similar to the one we are using to constrain our particular models. Although correct as a consistence test, it may be seen as a double counting of information. On the other hand, using the theoretical value for the acoustic scale would imply taking for granted that it coincides with the BAO scale and that the redshift of baryon-radiation drag is that obtained from CMB with the $\Lambda$CDM model. In order to avoid these assumptions and, at the same time, to verify the robustness of our results, we have also performed an analysis of the $H_0$ + SN + BAO + $\ell_{\mathrm{A}}$ set of observations with a flat prior on $r_s$, whose results are shown in Table \ref{tab:flat}. The values obtained for $r_s$ are in agreement with the Gaussian prior previously used. The evidence in favour of the interacting models persists, albeit inconclusive, with $|\ln B(\mathbf{D})| < 1$. For the GCG parameter we obtain the $2 \sigma$ interval $\alpha = -0.24^{+0.35}_{-0.30}$.

\begin{table*}
	\centering
	\begin{tabular}{c c c c}
	\hline
	\hline
	Data    			& Model Favoured	& $| \ln \mathrm{B(\textbf{D})} |$& 	Jeffreys' Scale 		\\ \hline \\
	OHD+SN+BAO+$\ell_{\mathrm{A}}$	&		GCG		&	1.362 $\pm$ 0.513		& 	Positive Evidence	\\ [1ex]
	$H_0$+SN+BAO+$\ell_{\mathrm{A}}$&		GCG		&	3.314 $\pm$ 0.513		&	Moderate Evidence	\\ [1ex]
	\hline \\
	OHD+SN+BAO+$\ell_{\mathrm{A}}$	& $\Lambda$CDM	&	1.247 $\pm$ 0.514		&	Positive Evidence	\\ [1ex]
	$H_0$+SN+BAO+$\ell_{\mathrm{A}}$& $\Lambda(t)$CDM	&	1.627 $\pm$ 0.518		&	Positive Evidence	\\ [1ex]
	\hline
	\hline
	\end{tabular}
	\caption{Results of our Bayesian model selection analysis.}
	\label{tab:models}
	\end{table*}


\section*{Acknowledgements}

TF thanks the financial support from CAPES/Brazil. SC is partially supported by CNPq (grant no. 309792/2014-2). JSA acknowledges support from CNPq (grants no. 310790/2014-0 and 400471/2014-0) and FAPERJ (grant no. 204282).



\bibliography{biblio}

\label{lastpage}
\end{document}